\documentclass{revtex4}
\usepackage{graphicx}
\usepackage{amsmath}
\usepackage{amssymb}
\usepackage{bbm}
\usepackage{}

\begin{document}

\title{Dynamical dark energy can amplify the expansion rate of the Universe }

\author{Hai-Chao Zhang}
\email{zhanghc@siom.ac.cn}

\affiliation{Shanghai Institute of Optics and Fine Mechanics, Chinese Academy of Sciences, Shanghai 201800, China }

\date{\today}

\begin{abstract}
By adding a matter-coupled dark energy field to Einstein's General Relativity (GR), this paper proves that the dynamical dark energy field can change the frequency of photons from distant galaxies as well as from background radiation of remote Universe. Therefore, when the observed frequency-shift of the photons is entirely attributed to the temporal variation of the cosmic scale factor, the calculated expansion rate of the Universe will be slightly greater than its actual value. The predicted values of the temperature of the cosmic blackbody radiation in the past (future) of the Universe are slightly larger (gradually smaller and smaller) than those in the standard cosmology. Since the blackbody radiation becomes the present cosmic microwave background (CMB) and its present-day temperature is directly estimated according to the Planck's law of blackbody radiation, the measured value of the CMB temperature is independent of whether to consider the scalar field or not.
\end{abstract}

\maketitle

\emph{Introduction.}$-$More and more observed evidences show that there exists a peculiar form of energy in nature \cite{planck,DaEnergy,z27,z28}. The characteristics of this kind of energy are completely different from matter (including ordinary matter, dark matter and electromagnetic field), and are also very different from gravitational energy. The so-called dark energy is currently described by the cosmological constant $\Lambda$ in the standard $\Lambda$-cold-dark-matter ($\Lambda$CDM) cosmological model \cite{z26,z100}. The value of $\Lambda$ in energy scale is equal to $\Lambda_{\mathrm{E}} \approx 2.24 \,\rm{meV}$ \cite{planck}, neither a `natural' value of zero \cite{z2,rrc1} nor another `natural' value $M_{\mathrm{Pl}} \approx 2.4\times 10^{18}\, \rm{GeV} $ of the Planck energy \cite{z29,z30,z31,z32}. The alternative scheme for describing the dark energy is to introduce a scalar field with self-interaction potential density (SPD) into GR\cite{rrc1,z29,z30}. In fact, as early as the 1960s, in order to be compatible with the Mach principle (MP), Brans and Dicke demonstrated that the relativistic gravitational theory should include a matter-coupled scalar field \cite{BD1,BD2}. Scalar fields are also introduced in the cosmic inflationary models to drive inflation of the early Universe \cite{KAO} so as to solve the flatness and horizon problems in the standard big-bang cosmology. The scalar field would couple to the other  forms of energy and then mediate a new force in nature \cite{Valerio,hcz2,DJP}.

In this paper, I will first prove that when dark energy is a dynamical scalar field and is conformally coupled with all other fields (except the gravitational field) in the same fashion, GR can maintain its traditional mathematical form. This is in line with Weinberg's theorem that a Lorentz invariant theory for spin-two-graviton must be GR \cite{wenb1,wenb2}. Based on the conformal coupling, I will then prove that the frequencies of the photons from the distant Universe can be shifted by the scalar field. The photons are described by the usual massless particles, differing from the extended theories of electromagnetism \cite{Sarracino,Alessandro} in which the massive photon frequency shift is discussed. The frequency-shift of the massless photons due to the scalar field would result in a false increase of the cosmic expansion rate when the additional redshift is also attributed to the Hubble parameter. The introduction technique of the coupling between the scalar field and other fields is similar to \cite{kaom,SC,z18,Calabrese}. However, I assume that the conformal coupling factor is: (i) the same (universal) function of the scalar field for all other fields (except for the space-time itself: gravitational field), so that the weak equivalence principle (WEP) in the presence of the scalar field is still valid not only for massive particles, but also for massless particles; (ii) a spontaneous symmetry-breaking function of the scalar field, so that the consequences of the scalar field appear only on both cosmic and smaller scales to satisfy the observed evidence of the accelerating expansion of the Universe and no observed evidence of existing a long-range scalar field. This direct coupling fashion leads to the disappearance of the physical concepts of Einstein frame (EF) and Jordan frame (JF) in our scheme. The frame concepts always appear in the usual scalar-tensor modified gravity theories \cite{Flanagan,jdb,hcz1,LUCa,P221204,AN,Anil,Onder,Bruck}.

With the improvement of astronomical measurement accuracy, the number of observable quantities related to methodology is also increasing, which offers ways to break the parametric degeneracy in a cosmic model. The parametric degeneracy refers to that it is difficult to distinguish the values of model parameters by the same set of fixed astronomical observations. Taking the Hubble constant$H_0$ as an example, various methodologies and techniques \cite{planck,4prdl03164,4prdl03141,4prdl03142,4prdl03144,4prdl03145,4prdl03147,4prdl03151,4prdl03162,4prdl03163} have been and will be developed to determine it with high accuracy. Current astronomical observations already show that the present-day value of the cosmic expansion rate seems to depend on methodology, called the Hubble constant tension \cite{zhang1123,z103,z109,4prdl03143,zhang1209}. This provides an opportunity to constrain the parameters of our model. In addition, since the scalar field can influence the value of the temperature of the CMB, the evolution of the temperature violates the traditional relation $T(z) = {T_{0}} \left( {1 + z} \right)$ of standard cosmology. The redshift range of the measurements for the redshift dependent temperature of the CMB has been extended to $z \sim 6.34$ \cite{4prdl03161,4prd202304011}. We can test the dynamical dark energy model with the data of the measurements \cite{4prdl03161,4prd202304011,4prd202304012,4prd202304013,4prd202304014,4prd202304015,4prd202304016,4prd202304017,4prd202304018}.

\emph{Setup.}$-$In order to maintain WEP \cite{hczhang4}, we require that various matter fields (including electromagnetic field) couple to the dark energy field $\phi$ via the same conformal fashion, i.e., ${\mathcal{L}_I} = \mathcal{B}\left( \phi  \right){\mathcal{L}_m}$, where ${\mathcal{L}_m}$ is the Lorents-invariant Lagrangian density for any kind of matter fields and $\mathcal{B}( \phi)$ is a dimensionless conformal coupling function of $\phi$, respectively. The dynamical dark energy field, matter fields, and the conformal interaction between the scalar and matter are minimally coupled to gravitational field, i.e., the total action is the sum as follows:
\begin{equation}
S =S_R+S_m+S_\phi+S_I= \int {{d^4}x\sqrt { - g} } \left( {\frac{1}{{16\pi G}}R+ {\mathcal{L}_m}+ {\mathcal{L}_\phi } +  \mathcal{B}\left( \phi  \right){\mathcal{L}_m}} \right),\label{3prdgr1}
\end{equation}
where $G$ is Newton's gravitational constant, $g$ is the determinant of the metric tensor $g_{\mu\nu}$, $R$ is the Ricci curvature scalar, and ${\mathcal{L}_\phi } =  - {g^{\mu \nu }}{\partial _\mu }\phi {\partial _\nu }\phi /2 - V(\phi )$ is the Lagrangian density for the scalar field with a scalar potential density $V(\phi )$ being the SPD. The metric tensor is related to the spacetime interval $ds$ by $ds^2=g_{\mu\nu}dx^\mu dx^\nu$, where the Einstein summation convention for repeated indices is used. The units $c=\hbar=1$ are adopted unless we emphasize the importance of the speed of light or the reduced Planck's constant.

In the presence of the scalar field, the Einstein's equations are obtained by varying the action (\ref{3prdgr1}) with respect to the metric as follows:
\begin{equation}
G_{\mu\nu}={8\pi G}\left( T^{(m)}_{\mu\nu}  +T^{(\phi)}_{\mu\nu}+  {\mathcal{B}\left( \phi  \right)} T^{(m)}_{\mu\nu}\right),\label{c4}
\end{equation}
where $G_{\mu\nu}$ is the Einstein tensor, and
\begin{equation}
{T_{\mu \nu }} =  - \frac{2}{{\sqrt { - g} }}\frac{{\partial \left( {\sqrt { - g} \mathcal{L}} \right)}}{{\partial {g^{\mu \nu }}}} \label{3prd1018}
\end{equation}
defines the energy-momentum tensor for any kind of energy, respectively. It is assumed that the coupling function does not depend explicitly on the metric and the Lagrangian density does not contain any derivatives of the metric.

The equation of motion for the scalar field in the presence of gravitational field is obtained by varying the action (\ref{3prdgr1}) with respect to $\phi$ as follows:
\begin{equation}
D^\mu D_\mu \phi=V_{,\phi} (\phi) -  {\mathcal{B}_{,\phi}\left( \phi  \right) }{\mathcal{L}_m},\label{3prd1n1}
\end{equation}
where $D_\mu$ is the covariant derivative with respect to the metric $g_{\mu\nu}$, $D^\mu$ is the contravariant operator defined by $D^\mu=g^{\mu\nu}D_\nu$, and the subscript $``,\phi"$ denotes a partial derivative of $\partial/\partial \phi$. Eq. (\ref{3prd1n1}) means that the scalar moves in an effective potential density (EPD) as follows:
\begin{equation}
{V_{{\text{eff}}}}\left( \phi  \right) = V\left( \phi  \right) - {\mathcal{L}_m}\mathcal{B}\left( \phi  \right).\label{tq8}
\end{equation}
One sees that the motion of the quintessence field is related to the Lagrangian density $ {\mathcal{L}_m}$ of matter rather than the trace $T_m$ of the energy-momentum tensor. Thus, although $T_m=0$ is always satisfied for electromagnetic field \cite{landau}, the motion of the scalar is directly influenced by electromagnetic field due to $ {\mathcal{L}_m}\neq 0$ in general. Of course, the value of the Lagrangian density for a plane electromagnetic wave indeed vanishes. Incidentally, we know that there are many Lagrangian densities of a kind of matter field that can give the same equation of motion for the matter field. However, the Lagrangian densities are also restricted by the requirement to give the correct relativistic energy-momentum tensor. Taking a specifical example, the Lagrangian for a free classical particle should not only give the correct equation of motion, but also the correct relativistic four-momentum whose time component is equal to the corresponding Hamiltonian and the space components are the relativistic three-dimensional momentum. If, in a sense of physics rather than mathematics, these Lagrangian densities for matter fields can still not be uniquely determined by all the above restrictions, the equation of motion for the scalar field shown as Eq. (\ref{3prd1n1}) will place considerable restrictions on the possible Lagrangian densities. Whether the choice of Lagrangian density is correct is ultimately determined by experiments.

From Eq. (\ref{c4}), it is evident that the total energy-momentum tensor of matter plus the scalar field must have a vanishing covariant divergence, i.e.,
\begin{equation}
\left( T^{(m)}_{\mu\nu} + {\mathcal{B}\left( \phi  \right)}T^{(m)}_{\mu\nu} +T^{(\phi)}_{\mu\nu}\right)_{;\nu}  = 0, \\ \label{3prdgr092923}
\end{equation}
where the subscript $``;\nu"$ denotes the covariant derivative $ D_{\nu}$. Both the energy-momentum tensor $T^{(\phi)}_{\mu\nu}$ of the scalar field and $T^{(m)}_{\mu\nu}$ of matter are no longer conserved respectively due to the conformal coupling between the scalar and matter.

We now discuss the equations of motion for a classical particle with rest mass $M$ in the presence of both the gravitational field and the dark energy field. Let us first introduce the four-momentum of the particle. We use a set of coordinate $x^\mu(\sigma)$ to describe a curve with $\sigma$ being a single parameter. The bare action for the classical particle is \cite{landau,3prl1122}
\begin{equation}
S_m = -M\int d\tau  =-M\int \frac{d \tau}{d\sigma}d\sigma, \label{3prd1115}
\end{equation}
where $\tau$ is the proper time of the particle. Since the moving speed of a particle cannot exceed the speed of light, the square of the spacetime interval between two events is always nonpositive under the $(-+++)$ metric convention, i.e., $ds^2 \leq 0$. Therefore, we introduce the proper time through $d\tau^2 = -ds^2$ so that the mass parameter $M$ is real. From Eq. (\ref{3prd1115}), the Lagrangian for the massive particle is $L=-M d\tau/d\sigma$, and then the Lagrangian density can be written as ${\mathcal{L}_m} = - M \delta \left( {{\mathbf{r}} - {{\mathbf{r}}(\sigma)}} \right) d\tau/d\sigma $, where $\delta \left( {{\mathbf{r}} - {{\mathbf{r}}(\sigma)}} \right)$ is Dirac $\delta$ function with spatial position vector ${\mathbf{r}}(\sigma)$ of the particle marking its spatial trajectory. The four-momentum $p_\mu$ of the classical particle can be defined as \cite{landau}
\begin{equation}
p_\mu \equiv \frac{{\partial S_m}}{{\partial {x^\mu }}}.\label{scicence1115309}
\end{equation}
Thus, by using the chain rule of differentiation to Eq. (\ref{3prd1115}) we obtain an important relation
\begin{equation}
p_\mu \frac{dx^{\mu}}{d\sigma} = -M \frac{d\tau}{d\sigma} \label{3prd11152}
\end{equation}
which is equivalent to the expression $p_\mu dx^{\mu}/ d\tau=-M $. This relation is superior to the usual mass-energy relation $p_\mu p^\mu=-M^2$ in calculation, where $p^\mu\equiv g^{\mu\nu}p_\nu=M dx^\mu/d\tau$ is the contravariant four-momentum of the particle.

Mathematically, the choice of a coordinate system in GR is not limited in principle. However, considering the comparison with experiments, such an appropriate physical coordinate system is often selected as far as possible, in which the time component $p^0$ can represent the particle's energy $\varepsilon$ measured by an observer who is at rest in the coordinates, i.e., $p^0=\varepsilon$, and the space components of $p^\mu$ correspond to the three-dimensional momentum $\mathbf{p}$ measured by the observer.

Based on the above definition of the four-momentum and the principle of least action \cite{hczhang4,landau}, the equations of motion for the classical particle can be obtained by varying the action (\ref{3prdgr1}) with respect to the path $x^\mu(\sigma)$ as follows:
\begin{equation}
{\left[ {\left( {1 + \mathcal{B}} \right){p^\mu }} \right]_{;\nu }}\frac{{d{x^\nu }}}{{d\tau}} =  - {M}\partial ^\mu \mathcal{B}, \label{scicence11092}
\end{equation}
or, equivalently,
\begin{equation}
{\left[ {\left( {1 + \mathcal{B}} \right){p^\mu }} \right]_{;\nu }}\frac{{d{x^\nu }}}{{d\sigma}} =  - {M}\partial ^\mu \mathcal{B} \frac{d\tau}{d\sigma} , \label{scicence11154}
\end{equation}
where $\partial ^\mu \equiv g^{\mu\nu}\partial_\nu$ with $\partial_\nu\equiv \partial/\partial x^\nu $. The particle no longer moves on a geodesic line since it couples to the dynamical dark energy field. When the coupling $\mathcal{B}( \phi)=0$, Eqs. (\ref{scicence11092}) and (\ref{scicence11154}) become the expected geodesic equations.

However, Eqs. (\ref{scicence11092}) cannot be used for the propagation of light signal because the denominator $d\tau$ on the left-hand of Eqs. (\ref{scicence11092}) equals to zero for light \cite{landau}. We have to use the single parameter $\sigma$ to describe the spacetime path of light ray. By inserting $d\tau=0$ into Eqs. (\ref{scicence11154}), the equations of motion for a light signal is obtained as follows:
\begin{equation}
{\left[ {\left( {1 + \mathcal{B}} \right){k^\mu }} \right]_{;\nu }}\frac{{d{x^\nu }}}{{d\sigma }} = 0, \label{scicence11093}
\end{equation}
where $k^\mu $ stands for the wave four-vector of the electromagnetic wavepacket, defined by the generalized de Broglie relation $p^\mu=\hbar k^\mu$ with $\hbar$ being the reduced Planck's constant. We see that, due to the scalar field, the magnitude of the tangent vector to the parameterized path of light signal will change, and then light signal does not fall on a geodesic line in a narrow sense. The geodesic line in the narrow sense refers to that, when a particle moves along the geodesic line, both the direction and the magnitude of the tangent vector to the parameterized path of the particle do not change. However, Eq. (\ref{scicence11093}) tells us that the four-momentum $k^\mu$ of the photon as well as the resultant four-momentum $\left( {1 + \mathcal{B}} \right){k^\mu }$ is indeed transported in parallel by the photon itself along the path curve. Thus, this path of the photon parameterized with $\sigma$ can be regarded as a geodesic curve in a more general sense, along which the direction of the four-momentum vector remains unchanged but its size can be changed. If we introduce a new mathematical curve parameter $\widetilde{\sigma}$ by the conformal transformation $d\widetilde{\sigma}=d\sigma/\left( {1 + \mathcal{B}} \right)$, where $\mathcal{B}$ is a function of $\sigma $ through the scalar field, then Eqs. (\ref{scicence11093}) become the conventional geodesic equations with respect to the new mathematical parameter $\widetilde{\sigma}$ in the above narrow sense. However, we should keep in mind that the measured physical quantities for photons are always expressed by the parameter $\sigma $ rather than by the rescaled parameter $\widetilde{\sigma}$.

Incidentally, even if the new rescaled curve parameter is used, the equations of motion for massive particles shown as Eqs. (\ref{scicence11092}) and (\ref{scicence11154}) still cannot become the conventional forms of the geodesic equations in the above narrow sense. No physical interaction can disappear globally by virtue of pure mathematical transformations. If the influence of a physical field on all matter fields can be eliminated by the same mathematical transformations, then this field cannot be introduced through measurable physical quantities, and there is no need to introduce it.

It is well known that the time component $k^0$ represents the wavepacket's frequency $\omega$, i.e., $k^0=\omega$, and the space components of $k^\mu$ correspond to the three-dimensional wave vector $\mathbf{k}$ of the wavepacket. Mathematically, the frequency $\omega$ represents an average value of the frequencies corresponding to the monochromatic components of the light wavepacket. The three-dimensional wave vector $\mathbf{k}$ represents an average wave vector of the expanded plane waves of the wavepacket. The relation of the frequency and the wave vector for light signal can be obtained by inserting $d\tau=0$ into Eq. (\ref{3prd11152}) as follows:
\begin{equation}
{k_\mu }\frac{{d{x^\mu }}}{{d\sigma }} = 0 \label{scicence11156}
\end{equation}
which is a generalized result of the usual relation $k^\mu k_\mu=0$ for monochromatic plane waves \cite{landau}. This generalized relation is superior to the usual one in calculation.

The curve parameter $\sigma$ must be introduced for the case of light signal since its proper time cannot be used to label its path. From a mathematical point of view, as an intermediate parameter, $\sigma$ is neither unique nor appears in the final solutions of the equations. In physics, the curve parameter could, but need not, be thought of the synchronous time for all of positions along the actual spatial trajectory of the light ray. Since the parameter appears in the equation of motion after all, it can also be said that $\sigma$ is determined by Eqs. (\ref{scicence11093}) and (\ref{scicence11156}).

Unlike the curve parameter $\sigma$, the mass parameter $M$ completely disappears in the equations of motion for light signal. It can only be seen as an auxiliary dimensional parameter \cite{3prl1122} in the action (\ref{3prd1115}) in obtaining Eqs. (\ref{scicence11093}) and (\ref{scicence11156}) for photons. These equations can also be obtained by inserting $M=0$ into Eqs. (\ref{scicence11154}) and (\ref{3prd11152}) of massive particles, meaning that photons are indeed zero rest-mass, but not zero energy particles. Due to the non-zero energy of photons, the motion of photons can be influenced by the scalar and gravitational fields, and vice versa.

\emph{A concrete form of conformal coupling.}$-$Since there is no observed evidence that a long-range scalar field exists, one can impose that the conformal coupling possesses a discrete $\mathbb{Z}_2$ spontaneously-broken symmetry \cite{LMK} and the SPD contains only $\phi^4$ term, i.e., \cite{hcz1}
\begin{subequations}\label{tq9}
\begin{eqnarray}
\mathcal{B}\left( \phi  \right) &=& \frac{1}{{4{M_1}^4}}{\left( {{\phi ^2} - {M_2}^2} \right)^2},\label{tq9a}\\
V(\phi ) &=& \frac{\lambda }{{\rm{4}}}{\phi ^4},\label{tq9b}
\end{eqnarray}
\end{subequations}
where ${M_1}$, ${M_2}$ and $\lambda $ are the three model parameters.

In cosmology, the matter content is regarded as a set of perfect fluids indexed by $i$, each with its energy density ${\rho _i}$ and pressure $P_i$. Since we consider a system of noninteracting particles, the Lagrangian density for a perfect fluid indexed by $i$ can be expressed as \cite{landau}
\begin{equation}
{\mathcal{L}_{mi}} =   - \sum\limits_C {\delta \left( {{\mathbf{r}} - {{\mathbf{r}}_C}} \right)\left( {{\varepsilon _C} - {{\mathbf{p}}_C} \cdot \frac{\partial \varepsilon_C}{\partial \mathbf{p}_C}} \right)}\equiv - \rho_i  + 3P_i ,\label{tq2211051}
\end{equation}
where $\varepsilon_C$ and $\mathbf{p}_C$ denote respectively the energy and the momentum of the particle $C$. The energy density $\rho_i$ and the pressure $P_i$ of the ideal fluid are obtained by averaging over all the particles in unit volume \cite{landau}. Obviously, for radiations and relativistic particles, $P_i =\rho_i/3$; for dust and cold dark matter, $P_i=0$. Thus, the total Lagrangian density of all perfect fluids in the Universe is,
\begin{equation}
{\mathcal{L}_m} = - \rho + 3P \equiv  -  \rho_{\mathrm{N}},\label{tq1105}
\end{equation}
where $\rho =  \sum \nolimits_i \rho_i$, $P  =  \sum \nolimits_i P_i$ and $\rho_\mathrm{N}\equiv \rho-3P$ denote the total energy density, the total pressure, and the effective nonrelativistic energy density of the total perfect fluids, respectively. Obviously, the effective nonrelativistic energy density $\rho_\mathrm{N}$ is in general smaller than the total energy density $\rho$. If all fluids contain only radiations and relativistic particles, then $\rho_\mathrm{N}=0$. If all fluids contain only dust and cold dark matter, then $\rho_\mathrm{N}=\rho$. Based on the cosmological constraint, it has been demonstrated that, after the era of the big bang nucleosynthesis (BBN) \cite{pbac,hcz1}, any quintessence field should and would sit stably at the minimum $\phi_{b}$ of the EPD. Substituting Eqs. (\ref{tq9}) and (\ref{tq1105}) into Eq. (\ref{tq8}), the minimum $\phi_{b}$ and the mass $m_b$ of the scalar field at the minimum are obtained, respectively, as follows:
\begin{subequations}\label{tq11}
\begin{eqnarray}
{\phi _{b }}^2 & = & \frac{{{ M_2}^2 \rho_\mathrm{N}}}{{\lambda{ M_1}^4 + \rho_\mathrm{N} }} ,\label{tq11a} \\
{m_{{b}}}^2 &  = & \frac{{2 {M_2}^2 \rho_\mathrm{N}}}{{{M_1}^4}}.\label{tq11b}
\end{eqnarray}
\end{subequations}
The mass of the quintessence strongly depends on the effective nonrelativistic matter density, leading to a short interaction range for a large matter density. However, the minimum of the EPD weakly depends on the effective nonrelativistic matter density when $\rho_\mathrm{N}  \gg \lambda {M_1}^4$, leading to a meaningful cosmological constant which will be shown by Eq. (\ref{3prde09304}) or equivalent Eq. (\ref{prl230122}).

\emph{Cosmological evolution.}$-$We consider a homogenous and isotropic cosmology with a scale factor $a(t)$ described by the line elements
\begin{equation}
d{s^2} =  - d{t^2} + {a^2}\left( t \right)\left[ {\frac{{d{r^2}}}{{1 - K{r^2}}} + {r^2}\left( {d{\theta ^2} + {{\sin }^2}\theta d{\varphi ^2}} \right)} \right],\label{3prdgr2}
\end{equation}
where $K=1$, $0$, or $-1$ correspond to closed, flat, or open spaces, respectively. The spaces contain several species of noninteracting perfect fluids of matter source. For a perfect fluid with Lagrangian density ${\mathcal{L}_{mi}}$ shown as Eq. (\ref{tq2211051}), the energy-momentum tensor is
\begin{equation}
T_{(mi)}^{\mu\nu}=(\rho_i+P_i)u^{\mu}u^{\nu}+P_i g^{\mu \nu},\label{3prd1001}
\end{equation}
where $u^\mu$ is the four-velocity for the macroscopic motion of an element of volume of the fluid. $u^\mu$ differs from the four-velocity for microscopic motion of the particles.

Using Eqs. (\ref{3prd1n1}), (\ref{3prdgr2}) and (\ref{3prd1001}), the equations of motion for the quintessence and the perfect fluids are
\begin{subequations}\label{3prde1003}
\begin{eqnarray}
\ddot \phi  + 3H\dot \phi  + \frac{1}{a^2(t)}{\nabla^2_{\rm{FRW}}}\phi  + {V_{,\phi }}\left( \phi  \right) + {\mathcal{B}_{,\phi }}\left( \phi  \right)\rho_\mathrm{N}  = 0, \label{3prde10031}\\
{{\dot \rho }_i} + 3H\left( {{\rho _i} + {P_i}} \right) = \frac{-3P_i}{{ {1 + \mathcal{B}} }}\frac{{d\mathcal{B}}}{{dt}},\label{3prde10032}
\end{eqnarray}
\end{subequations}
respectively, where the overdots denote derivatives with respective to time, $H = \dot a(t)/a(t)$ is the Hubble parameter and ${\nabla^2_{\rm{FRW}}}$ is the Laplacian operator \cite{3prdnote} corresponding to the Friedmann-Roberson-Walker (FRW) metric shown as Eq. (\ref{3prdgr2}). Suppose that the cosmic background consists of only two extreme types of matter: nonrelativistic particles $P_i=0$ and relativistic particles $P_i=\rho_i/3$, from Eq. (\ref{3prde10032}), we have
\begin{equation}
\dot{\rho}_\mathrm{N}=-3H {\rho}_\mathrm{N}. \label{prl1126}
\end{equation}

One sees that Eq. (\ref{3prde10031}) describes a damped (negative damped) oscillation of the quintessence for an expanding (a contracting) universe corresponding to $H>0$ ($H<0$). The rate of change of the minimum with respect to the time variable can be evaluated by Eqs. (\ref{tq11a}) and (\ref{prl1126}) as follows
\begin{equation}
\frac{{{{\dot \phi }_b}}}{{{\phi _b}}} =  - \frac{{3H}}{2}\frac{{\lambda {M_1}^4}}{{\lambda {M_1}^4 + {\rho _N}}}.\label{4prdl0311}
\end{equation}
In an expanding universe of $H>0$, we have ${\dot \phi }_b/{{\phi _b}} < 3H/2$, meaning that when the quintessence reaches the minimum, it will always follow the minimum adiabatically \cite{hcz1}. In a contracting universe of $H<0$, based on the positive feedback mechanism, even if the quintessence initially sits at the minimum, it will depart away the equilibrium value with time.

For our expanding Universe, with the time growing the quintessence eventually and already reaches the minimum $\phi_{b}$ of the EPD, i.e., ${V_{,\phi }} \left( \phi_{b}\right)+\rho_{\mathrm{N}} \mathcal{B}_{,\phi}\left( \phi_{b}  \right)= 0$. It then sits stably at the minimum since the rate of change of the minimum is sufficiently small compared with the cosmic expanding rate. The minimum $\phi_b$ indeed presents a dynamical equilibrium state for the quintessence in the homogeneous background medium of matter. Thus, when the equilibrium state is achieved, Eq. (\ref{3prde10031}) can be divided into the following equations:
\begin{subequations}\label{3prde1004}
\begin{eqnarray}
{V_{,\phi }}\left( {{\phi _b}} \right) + {\mathcal{B}_{,\phi }}\left( {{\phi _b}} \right)\rho_\mathrm{N} = 0, \label{3prde10041}\\
{\nabla^2_{\rm{FRW}}}{\phi _b} = 0,\label{3prde10042}\\
{{\ddot \phi }_b} + 3H{{\dot \phi }_b} = 0. \label{3prde10043}
\end{eqnarray}
\end{subequations}
These equations mean that, for the expanding Universe with the homogenous distribution of matter, the scalar field eventually and slowly evolves along the spatial-homogeneous equilibrium state.

In the equilibrium state of the quintessence, using Eqs. (\ref{c4}) and (\ref{3prdgr2}) we obtain the equations of cosmic evolution as follows:
\begin{subequations}\label{3prde09301}
\begin{eqnarray}
{H^2} &\equiv & {\left( {\frac{{\dot a}}{a}} \right)^2}  = \frac{{8\pi G}}{3}\left( {V\left( {{\phi _b}} \right) +{\rho}} \left(1 + {\mathcal{B}} \right) + \frac{1}{2}{{\dot \phi }_b}^2 \right) - \frac{K}{{{a^2}}},\label{3prde093011}\\
\frac{{\ddot a}}{a} &=& \frac{{4\pi G}}{3}\left( {2V\left( {{\phi _b}} \right) - \left(1 + {\mathcal{B}} \right)\left( {{\rho } + 3{P}} \right)} - 2{{\dot \phi }_b}^2 \right),\label{3prde093012}
\end{eqnarray}
\end{subequations}
where $\mathcal{B}\equiv\mathcal{B}\left( \phi_b  \right)$ specifically refers to the conformal coupling at the equilibrium state $ \phi_b $, $a\equiv a(t)$ is the scale factor, and $\rho = \rho_N+3P$ (see Eq. (\ref{tq1105})) denotes the cosmic total energy density in the absence of the scalar field and in terms of $\rho_N$ and $P$, so as to make the formula concise. The conformal coupling and the self-interaction potential density can be obtained by inserting Eq. (\ref{tq11a}) into Eqs. (\ref{tq9a}) and (\ref{tq9b}), respectively, as follows
\begin{subequations}\label{4prde20230311}
\begin{eqnarray}
\mathcal{B}\left( {{\phi _{b}}} \right) & = & \frac{1}{4}{\left( {\frac{{\lambda {M_1}^2{M_2}^2}}{{\lambda {M_1}^4 + \rho_\mathrm{N} }}} \right)^2}, \label{3prd10061}\\
V({\phi _b}) & = &\frac{\lambda }{4}{\left( {\frac{{{M_2}^2{\rho _{\text{N}}}}}{{\lambda {M_1}^4 + {\rho _{\text{N}}}}}} \right)^2}.\label{4prde202303112}
\end{eqnarray}
\end{subequations}
Thus, if the nonrelativistic matter density $\rho_\mathrm{N}$ is large enough, the value of the conformal coupling $\mathcal{B}\left( {{\phi _{b}}} \right)$ is far less than 1, and then the coupling function in Eqs. (\ref{3prde09301}) can be ignored. One can see that, roughly, the difference of $\rho -3P$ affects the scalar field, while the sum  $\rho +3P$ affects the gravitational field. Thus, for the same value of $\rho$, the source of a relativistic fluid has the largest gravitational effect, but no effect on the scalar field. By comparing Eqs. (\ref{3prde09301}) with the $\Lambda$CDM model \cite{z26,z100}, one can see that the value $V \left( {{\phi _b}} \right)$ of the SPD at the equilibrium state is similar to the cosmological constant in the $\Lambda$CDM model \cite{note1}. From Eq. (\ref{4prde202303112}), we can see that, as long as the parameters' combination $\lambda M_1^4 $ is smaller than the current cosmic density, the value of the SPD is very close to a constant $\lambda {M_2} ^4/4 $. Since the density of matter is quite large in the remote past of the expanding Universe, the limiting value $\lambda {M_2}^4/4$ multiplied by the factor $8\pi G$ can be defined as the cosmological constant $\Lambda$ at that time \cite{note1}, i.e.,
\begin{equation}\label{prl230122}
\Lambda =2\pi \lambda G {M_2}^4.
\end{equation}
Because the cosmological constant in energy scale is often used in literature, we define it by the limit as follows:
\begin{equation}
{\Lambda _{\text{E}}} = \mathop {\lim }\limits_{{\rho _N} \to \infty } {V^{1/4}}\left( {{\phi _b}} \right) = {\left( {\frac{\lambda }{4}} \right)^{1/4}}{M_2}.\label{3prde09304}
\end{equation}
The $\Lambda$CDM model can be regarded as an approximate form of the dynamical dark energy model in the case of large nonrelativistic matter densities of the Universe.

The orders of the model parameters are constrained theoretically and experimentally as follows \cite{planck,auw,DJP,hcz2}: $\lambda\sim\mathcal{O}(1) $, $M_2 \sim \mathcal{O}(\Lambda_E)$, $ M_1/M_2 \sim \mathcal{O}(1)$, and the current cosmic density $\sim \mathcal{O}({\Lambda_\mathrm{E}}^4)$. Noticing Eqs. (\ref{tq11a}) and (\ref{prl1126}), and the definition of the reduced Planck energy ${M_{\mathrm{Pl}}}=1/\sqrt{8\pi G}$, one can see that the kinetic energy density (KED) terms of ${{\dot \phi }_b}^2 /2$ and $2{{\dot \phi }_b}^2$ in Eqs. (\ref{3prde09301}) can be safely neglected. In Eqs. (\ref{3prde09301}), the ratios of the KED terms to the SPD terms are estimated to be far smaller than the ratio of ${(\Lambda_\mathrm{E }/M_{\mathrm{Pl}})}^2 \sim \mathcal{O}(10^{-60})$. Incidentally, only the combinations $\lambda M_1^4 $ and $\lambda M_2^4 $ appear in the expressions of $\mathcal{B}\left( \phi_b  \right)$ and $V\left( {{\phi _b}} \right)$. Thus, the equations of cosmic evolution shown by Eqs. (\ref{3prde09301}) are only related to these combinations. This results in that the separate values of $\lambda$, $ M_1 $ and $ M_2 $ cannot be determined through astronomical observations based on the homogenous and isotropic cosmology.

\emph{Frequency-shift of light signal.}$-$The information of the cosmic evolution is often obtained by measuring the electromagnetic radiations from distant galaxies and the CMB from the early Universe. Consider a test signal which is emitted from atoms in a distant comoving galaxy at the late Universe. The light signal passes through the cosmic perfect fluids for a long journey and is eventually measured by a comoving local observer on our Earth. The original frequency of the signal is naturally equal to that of photons emitted by the same kind of atoms in the laboratory on the Earth. By comparing the received frequency of the light signal with the reference frequency in the laboratory, one can determine the frequency shift of the light signal. Using Eqs. (\ref{scicence11093}) and (\ref{scicence11156}) under the FRW metric shown as Eq. (\ref{3prdgr2}), the curve parameter $\sigma$ is determined to be the time variable $t$ and the evolution equation of the frequency $\omega $ of the light signal is determined as follows (see appendix \ref{appendixA}):
\begin{equation}
\frac{{d\omega }}{{dt}} = - \left( { H + \frac{1}{{ {1 + \mathcal{B}} }}\frac{{d\mathcal{B}}}{{dt}}} \right)\omega \label{3prl1123}
\end{equation}
with $H\equiv \dot a(t)/a(t)$ being the Hubble parameter. Eq. (\ref{3prl1123}) means that
\begin{equation}
\left( {1 + \mathcal{B}} \right)\omega a(t) = \mathrm{constant}. \label{3prl11232}
\end{equation}
When the coupling $ \mathcal{B}= 0$, the above equation becomes the old expression $\omega a(t) = \mathrm{constant}$ which give us the well-known redshift for the expanding Universe. However, from Eq. (\ref{3prl1123}), one sees that the redshift of a light signal comes from two factors: the Hubble expansion and the temporal variation of the conformal coupling. Thus, if the observed frequency-shift of the light signal is only attributed to the expansion rate of the Universe, the calculated apparent value $H_{{\text{app}}}$ of the expansion rate will be larger than the actual value $H$ of the expansion rate. Due to $d\mathcal{B}/dt = \partial\mathcal{B}/\partial t$ in the spatial homogenous case, from Eq. (\ref{3prl1123}) we have
\begin{equation}
{H_{{\text{app}}}} = H + \frac{1}{{1 + \mathcal{B}\left( \phi_b  \right)}}\frac{{\partial \mathcal{B}\left( \phi_b  \right)}}{{\partial t}}. \label{3prd100522}
\end{equation}
Noticing Eq. (\ref{prl1126}), we can rewrite Eq. (\ref{3prd100522}) as follows:
\begin{equation}
{H_{{\text{app}}}} = H\left( {1 - \frac{{3\rho_\mathrm{N}}}{{1 + \mathcal{B}\left( {{\phi _{{b}}}} \right)}}\frac{{\partial \mathcal{B}\left( {{\phi _{{b}}}} \right)}}{{\partial \rho_\mathrm{N} }}} \right), \label{3prd10062}
\end{equation}
where the conformal coupling $\mathcal{B}\left( {{\phi _{{b}}}} \right)$ is related to the non-relativistic matter density $\rho_\mathrm{N}$ of the Universe through Eq. (\ref{3prd10061}).

\emph{$\mathcal{B}$-dependence and H-dependence of observable quantities.}$-$The Hubble parameter dependence and the coupling factor dependence are not the same for different observable quantities. For example, if the Universe expands adiabatically, then the frequency $\omega$ of photons in the CMB varies, but their number population $N(\omega)$ remains constant. Hence the temperature $T(t)$ of the CMB can be deduced by its spectral distribution according to the Plank's law under the adiabatic expansion, i.e.,
\begin{subequations}\label{4prdl3171}
\begin{eqnarray}
N(\omega (t)) & = & \frac{1}{{{e^{\hbar \omega (t)/{k_B}T(t)}} - 1}}, \label{4prdl3172}\\
N({\omega _0}) & = & \frac{1}{{{e^{\hbar {\omega _0}/{k_B}{T_0}}} - 1}},\label{4prdl3173}\\
N(\omega (t))& = & N({\omega _0}), \label{4prdl3174}
\end{eqnarray}
\end{subequations}
where $k_B$ is Boltzmann constant, $\omega _0$  and $T_0$ is the present-day values of frequency and temperature of the CMB measured on the Earth. Combining Eqs. (\ref{4prdl3171}) with Eq. (\ref{3prl1123}), the evolution equation of the temperature $T(t) $ of the CMB is obtain as follows
\begin{equation}
 \frac{{dT(t)}}{{dt}} =  - \left( {H + \frac{1}{{1 + \mathcal{B}}}\frac{{d\mathcal{B}}}{{dt}}} \right)T(t), \label{4prdl3177}
\end{equation}
which means that
\begin{equation}
\left( {1 + \mathcal{B}} \right)T(t) a(t) = \left( {1 + \mathcal{B}_0} \right)T_0 a_0, \label{4prdl3178}
\end{equation}
where $\mathcal{B}_0$ denotes the present-day value of the conformal coupling factor. When the coupling $ \mathcal{B}= 0$, the above equation becomes the old expression $T(t) a(t) = T_0 a_0$.

In cosmology, the cosmic time $t$ is customarily represented by the cosmological redshift $z$ defined through the ratio of the current cosmic scale $a_0$ to the cosmic scale $a(t)$ at time $t$ as follows
\begin{equation}
1 + z = \frac{{{a_0}}}{{a(t)}}. \label{4prdl3181}
\end{equation}
Hence inserting Eq. (\ref{4prdl3181}) into Eq. (\ref{4prdl3178}), we see that the CMB temperature will evolve as
\begin{equation}
 T(z) = {T_0} \left( {1 + z} \right)\frac{{ {1 + {\mathcal{B}_0}} }}{{ {1 + \mathcal{B}(z)} }}, \label{4prdl3182}
\end{equation}
which differs from the old expression
\begin{equation}
T'(z) = {T_0} \left( {1 + z} \right) \label{4prdl0428}
\end{equation}
in standard cosmology. Incidentally, we have the evolution of the frequency of a light signal with the redshit $z$ as
\begin{equation}
\omega (z) = {\omega _0}\left( {1 + z} \right)\frac{{1 + {\mathcal{B}_0}}}{{1 + \mathcal{B}(z)}}, \label{4prd0328e}
\end{equation}
corresponding to Eq. (\ref{3prl1123}). It should be emphasized that the meaning of redshift defined by Eq. (\ref{4prdl3181}) is different from that of its original definition in physics of redshift $z_f$ defined by
\begin{equation}
1 + z_f =\frac{\omega (t) }{\omega _0}. \label{4prdl04282}
\end{equation}

If light signal is emitted from atoms in a distant galaxy, its original frequency $\omega (t)$ should be equal to the frequency of photons emitted from the same kind of atoms in laboratory on the Earth, while ${\omega _0}$ is the measured value of the signal frequency on the Earth. If light signal comes from the cosmic blackbody radiation, based on the calculation of nucleosynthesis, the maximum in the energy spectrum of the blackbody radiation is about 1 MeV at the time $\sim$ 200 sec after the Big Bang \cite{4prd202304634}, corresponding to the temperature $\sim 10^9$ K. With the Universe expanding, the blackbody radiation becomes the present-day CMB with temperature estimated to be $\sim$ 5 K \cite{4prd202304635}, very close to the accurately measured value of $2.7255\pm 0.0006$ K \cite{4prdl03161}.

Since the number population density $n(t)$ and the energy density $\rho(t)$ of the CMB are defined as follows:
\begin{subequations}\label{4prdl31711}
\begin{eqnarray}
n(t) &=& \frac{{{n_0}{a_0}^3}}{{a{{(t)}^3}}}, \label{4prdl31712}\\
\rho (t) &=& n(t)\omega (t),\label{4prdl31713}
\end{eqnarray}
\end{subequations}
together with Eq. (\ref{3prl1123}), we have evolution equations for number density $n(t)$ and the energy density respectively as follows:
\begin{subequations}\label{4prdl31714}
\begin{eqnarray}
\frac{{dn(t)}}{{dt}}& = & - 3Hn(t) , \label{4prdl31715}\\
\frac{{d\rho (t)}}{{dt}} &= & - \left( {4H + \frac{1}{{1 + \mathcal{B}}}\frac{{d\mathcal{B}}}{{dt}}} \right)\rho (t) .\label{4prdl31716}
\end{eqnarray}
\end{subequations}

\emph{Explaining the Hubble constant tension.}$-$We have seen that, from Eqs.(\ref{4prdl31714}) and (\ref{3prl1123}), if the scalar field exists and its influence is ignored in measurements, different methodologies will give different estimates for the same parameter (such as the Hubble parameter). For the expanding homogenous Universe, the value of Hubble parameter will be amplified by the scalar field in measurements based solely on the redshift of light frequency, but will not be affected in the purely number density measurements. Due to the frequent mixing of these two mechanisms in practical methodologies, the estimated values of Hubble parameters are distributed between the two extremes mentioned above.

As a first step, it is reasonable to assume that the minimum value is close to the real value of the cosmic expansion rate, while the larger value is far away from it. Fortunately, various methodologies and techniques \cite{planck,4prdl03164,4prdl03141,4prdl03142,4prdl03144,4prdl03145,4prdl03147,4prdl03151,4prdl03162,4prdl03163} have been developed with high accuracy. It seems that the present-day value of the cosmic expansion rate depends on methodology, known as the Hubble constant tension \cite{zhang1123,z103,z109,4prdl03143,zhang1209}: the difference between values of the Hubble constant measured by the Planck experiment, $H_0 = 67.36 {^{+0.54}_{-0.54}} \,\rm{km\,s^{-1} \,Mpc^{-1}}$ \cite{planck}, and a local expansion rate measurement of the Hubble constant made using Cepheids and type Ia supernovae now $H_0 = 73.04 {^{+1.04}_{-1.04}} \,\rm{km\,s^{-1} \,Mpc^{-1}}$ \cite{4prdl03164}. There are many measurements of $H_0$ that lie between 67.36 and 73.04 $\rm{km\,s^{-1} \,Mpc^{-1}}$, with most of them being clustered around 68-70 $\rm{km\,s^{-1} \,Mpc^{-1}}$ \cite{4prdl03164,4prdl03141,4prdl03142,4prdl03144,4prdl03145,4prdl03147,4prdl03151,4prdl03162,4prdl03163}, and most of them being mutually consistent.

\begin{figure}
\centering
\includegraphics[width=250pt]{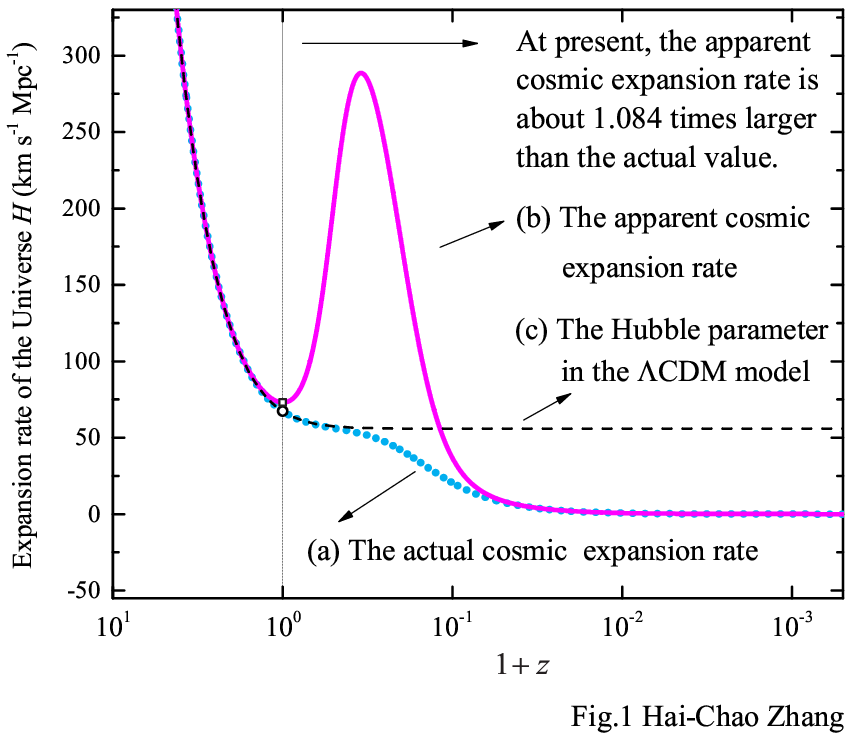}
\caption{Cosmic expansion rates vary with cosmic time denoted by cosmological redshift. (a) The actual expansion rate (dotted curve) is calculated by Eq. (\ref{3prde093011}). The following values of the model parameters and the cosmic parameters are used to satisfy the constraints of cosmology \cite{planck,zhang1123}, the fifth-force experiment \cite{DJP,hcz2} and the theory of quantum stability \cite{auw}: $\lambda=1/6$, $M_1=M_2/6.0816$, $M_2$ = 4.97251 meV, and $\rho_{N0}=2.65072\times 10^{-27} \,\rm{kg\,m^{-3}}$. The cosmic matter density is not a model parameter in our setup. These parameter-values are selected carefully so as to obtain the actual Hubble constant $H_0=67.36\,\rm{km\,s^{-1} \,Mpc^{-1}}$ (hollow circle, corresponding to the fitted value with the $\Lambda$CDM model from the distributions of the CMB \cite{planck}) and the flat space of $K=0$. If the space is closed (open)), the spatial curvature parameter $K=1$ ($K=-1$) can be achieved by slightly increasing (decreasing) the above critical density of $\rho_{N0}=2.65072\times 10^{-27} \,\rm{kg\,m^{-3}}$. (b) The apparent expansion rate (solid curve) is calculated by Eq. (\ref{3prd10062}), meaning that the redshift is attributed entirely to the cosmic expansion rate. The present-day value of the apparent expansion rate corresponds to the inferred value of the Hubble constant deduced from the late Universe \cite{zhang1123}, i.e., $H_{\mathrm{app0}}=73.04\,\rm{km\,s^{-1} \, Mpc^{-1}}$(hollow square). (c) The Hubble parameter in the $\Lambda$CDM model (dashed curve). By letting $V(\phi_b)=\Lambda_{\mathrm{E}}^4$ and $\mathcal{B}(\phi_b)={{\dot \phi }_b}^2=K=0$ in Eqs. (\ref{3prde09301}), the equations of cosmic evolution in the flat $\Lambda$CDM model can be obtained. There is no concept of apparent expansion rate in the $\Lambda$CDM model. }\label{figure1}
\end{figure}

We can, but we don't have to, choose a moderate value from the data set of the smaller values as the true value of the Hubble constant, e.g, $H_0 = 67.36 \,\rm{km\,s^{-1} \,Mpc^{-1}}$ \cite{planck} and choose another moderate value from the data set of the larger values as the amplified value of the Hubble constant, e.g, $H_{\mathrm{app0}} = 73.04 \,\rm{km\,s^{-1} \,Mpc^{-1}}$ \cite{4prdl03164}, then the evolutions of the Hubble parameter and the amplified Hubble parameter over time can be plotted in Fig. \ref{figure1}.

Figure \ref{figure1} shows that the actual and apparent expansion rates of the Universe as functions of redshift. In the remote past of the Universe, the apparent value $H_{{\text{app}}}$ is almost equal to the actual Hubble parameter $H$ due to $\mathcal{B}\left( \phi_b  \right)$ approaching zero for very large cosmic density. In the future, the apparent value $H_{{\text{app}}}$ will gradually become larger than the actual value $H$, and then will tend to reach the actual value $H$ again with the time moving forward due to $\mathcal{B}\left( \phi_b  \right)$ approaching a constant for extreme thin cosmic density. The model parameters ($\lambda$, $M_1$ and $M_2$) are independent of the time we live, but the density of cosmic matter varies over time. If we live in the far past or future of the Universe when the apparent value $H_{{\text{app}}}$ is almost equal to the actual Hubble parameter $H$, i.e., there is no ``Hubble tension", then the parametric degeneracy of the dynamical dark energy model cannot be broken by comparing $H_{{\text{app}}}$ with $H$.

\begin{figure}
\centering
\includegraphics[width=250pt]{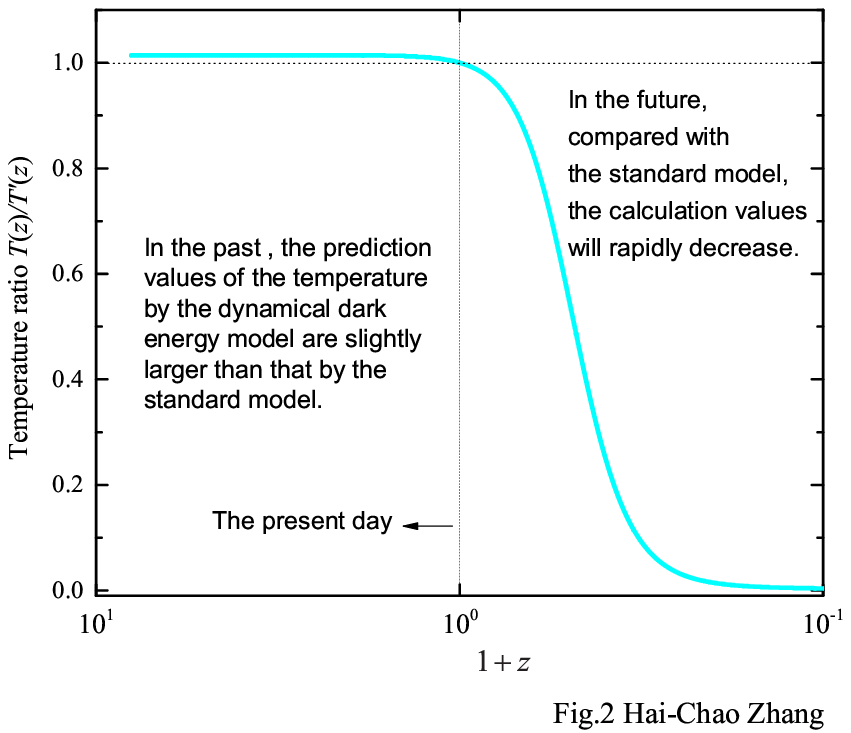}
\caption{The ratio between the temperature values in the dynamical dark energy model and that in the standard model versus redshift $z$. The values of the model parameters are the same as those used in Fig. \ref{figure1}. }\label{figure2}
\end{figure}

\emph{The effect of the scalar field to the CMB.}\cite{referree0429}$-$From Eq. (\ref{4prdl3182}), the value of the temperature of the CMB can be influenced by the scalar field. In order to compare the dynamical dark energy model with the standard cosmology in describing the temperature evolution of the CMB, we introduce a temperature ratio of $T(z)/T'(z)$ through Eqs. (\ref{4prdl3182}) and (\ref{4prdl0428}) as follows:
\begin{equation}
\frac{T(z)}{T'(z)} = \frac{{ {1 + {\mathcal{B}_0}} }}{{ {1 + \mathcal{B}(z)} }}. \label{4prd04301}
\end{equation}
The evolution of the ratio with redshift $z$ is shown in Fig. \ref{figure2}.

From figure \ref{figure2}, one sees that the calculated values of the CMB temperature by the dynamical dark energy model are slightly larger than that by the standard model in the past of the Universe. But in the future, compared with the standard model, the calculation values will be smaller and smaller.

\begin{figure}
\centering
\includegraphics[width=250pt]{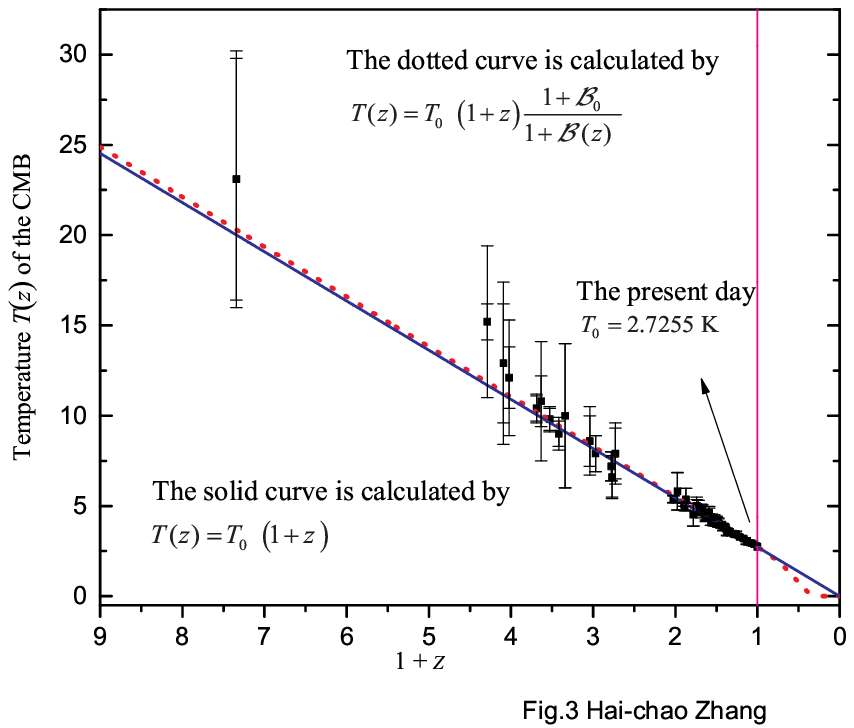}
\caption{The temperature evolution of the CMB for $H_0 = 67.36 \,\rm{km\,s^{-1} \,Mpc^{-1}}$. The dotted curve is calculated by Eq. (\ref{4prdl3182}) in the dynamical dark energy model. The solid curve is calculated by the relation $T'(z) = {T_{0}} \left( {1 + z} \right)$ in standard cosmology. The solid squares mark the measurement data listed in \cite{4prdl03161}. The values of the model parameters are the same as those used in Fig. \ref{figure1}. }\label{figure3}
\end{figure}

Since the redshift range of the measurements for the redshift dependent temperature of the CMB has been extended to $z \sim 6.34$ \cite{4prdl03161,4prd202304011}, we can compare the data of the measurements \cite{4prdl03161,4prd202304011,4prd202304012,4prd202304013,4prd202304014,4prd202304015,4prd202304016,4prd202304017,4prd202304018} with the calculated values by the dynamical dark energy model. The evolution of the CMB temperature
is shown in Fig. \ref{figure3}

One can see that, in order to reach the current temperature value of $2.7255$ K, the temperature of the CMB at non-zero-shifts should be higher than that predicted by the standard cosmology so that the CMB can transfer energy to the scalar field. However, from Eqs. (\ref{3prd10061}) and  (\ref{3prl1123}), the transfer process mainly occurs when the density of cosmic matter fluid becomes thinner. For dense matter fluids, the value of the conformal coupling $\mathcal{B}\left( {{\phi _{b}}} \right)$ is far less than 1, and then the amount of the transfer energy can be safely neglected. In the future, the cosmic density will becomes smaller and smaller, resulting in that the conformal coupling $\mathcal{B}\left( {{\phi _{b}}} \right)$ will become larger and larger. This will cause a rapid decrease in CMB temperature.

\begin{figure}
\centering
\includegraphics[width=250pt]{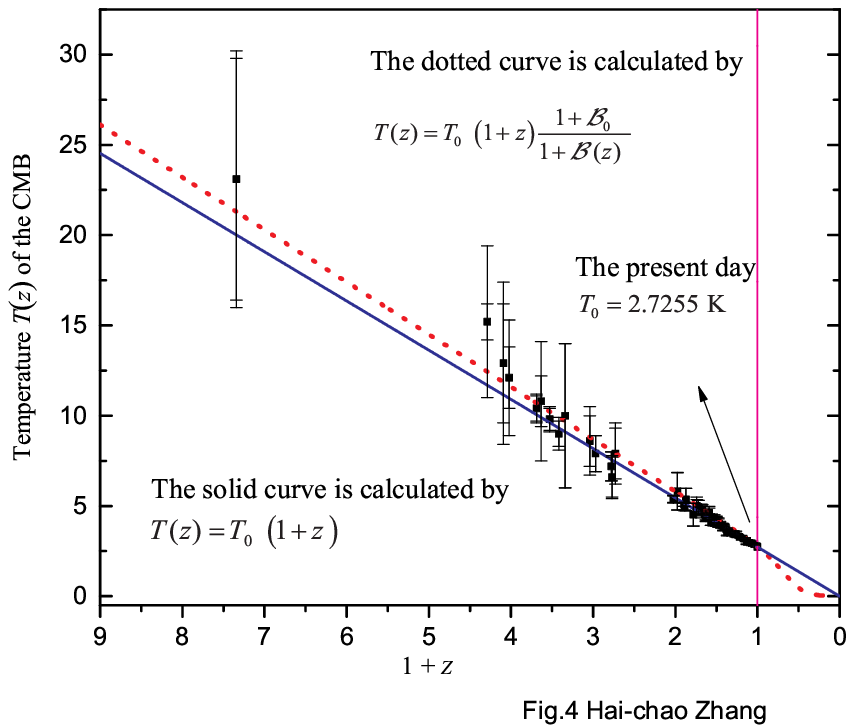}
\caption{The temperature evolution of the CMB for $H_0 = 54.4\,\rm{km\,s^{-1} \,Mpc^{-1}}$. The dotted curve is calculated by Eq. (\ref{4prdl3182}) in the dynamical dark energy model. The solid curve is calculated by the relation $T'(z) = {T_{0}} \left( {1 + z} \right)$ in standard cosmology. The solid squares mark the measurement data listed in \cite{4prdl03161}. In the case of $K=0$, the values of the model parameters are used in the calculation as follows: $\lambda=1/6$, $M_1=M_2/3.0586$, $M_2$ = 4.26149 meV, and $\rho_{N0}=2.54319\times 10^{-27} \,\rm{kg\,m^{-3}}$. }\label{figure4}
\end{figure}

According to figure \ref{figure3}, we can roughly say that our model can recover the standard cosmology in the past of the Universe. However, both calculated curves are far away from the measured data points at the larger redshifts. In fact, since the present-day value of the CMB temperature is definitely a constant, it is impossible to fit the measured data using the standard relation of $T'(z) = {T_0} \left( {1 + z} \right)$. We can use Eq. (\ref{4prdl3182}) to fit the data points very well by assuming $\rho_N=\rho_{N0}(1+z)^3$ in the pressureless case $P=0$ without any other constraints. However, for the best fitting result, the parameter ratio of $M_2/M_1$ is smaller than the critical value of $12^{1/4}=1.861$ \cite{20230505}. The critical value of $M_2/M_1$ corresponds to the intersecting condition of the first two terms (as functions of $z$) on the right-hand side of Eq. (\ref{3prde093012}). When the ratio of $M_2/M_1$ is smaller than $12^{1/4}$, the model cannot provide enough dark energy to drive the accelerated expansion of the Universe at the late time. Therefore, the experimental data points should be fitted under the constraints of the ratio $M_2/M_1>12^{1/4}$ and the transition redshift of deceleration-acceleration $z_{\texttt{tran}}\sim 0.5$. If we still maintain the flat space of $K=0$ and $H_{\mathrm{app0}}=73.04\,\rm{km\,s^{-1} \, Mpc^{-1}}$ as described above, a much smaller value of Hubble rate should be introduced so as to obtain large values of temperature for the larger redshifts as shown by the experimental data. The smaller the introduced Hubble constant, the higher the CMB temperature value at a large redshift. Based on the data of the measurements \cite{4prdl03161,4prd202304011,4prd202304012,4prd202304013,4prd202304014,4prd202304015,4prd202304016,4prd202304017,4prd202304018}, the value of $H_0$ is deduced to be about $54.4\,\rm{km\,s^{-1} \,Mpc^{-1}}$ from the fitting curve under the cosmic constraints above. The calculation curve of the CMB temperature corresponding to $H_0=54.4\,\rm{km\,s^{-1} \,Mpc^{-1}}$ is shown in Fig. \ref{figure4}.

Consequently, if the scalar field exists and can be described by our model, the present-day expansion rate of the Universe may be smaller than $67.36\,\rm{km\,s^{-1} \,Mpc^{-1}}$ \cite{planck}. The estimated value of $H_0=54.4\,\rm{km\,s^{-1} \,Mpc^{-1}}$ is exactly equal to the value reported in the literature \cite{z114}. However, the value of $H_0$ is deduced by fitting the CMB temperature data in the case of flat space with the dynamical dark energy model. In the literature \cite{z114}, the same value of $H_0$ is obtained by the data of the Planck experiment with the $\Lambda$CDM in the case of the closed space. In fact, one can find that the same value of $H_0$ can also be deduced by fitting the CMB temperature data in the case of closed space with the dynamical dark energy model. This implies that the temperature evolution of the CMB is independent of the spatial curvature of the Universe, but it is sensitive to the cosmic expansion rate in the dynamical dark energy model.

It should be pointed out that for large redshifts, there are not many experimental data points of the CMB temperature and there are also significant measurement errors in the existing measurement data. Therefore, it is difficult to accurately determine the fitting parameters. Further dense and accurate measurements of CMB temperature are needed in the case of large redshift.

\emph{Conclusions.}$-$The scalar field for dark energy is introduced through its self-interaction potential and the conformal interaction potential with matter. The ratio of the interaction Lagrangian density and the bare Lagrangian density of matter does not depend on the nature of matter, it is a universal and spontaneous symmetry-breaking function of the scalar field. The dynamical property of the dark energy field can cause a new redshift for the frequency of a light signal from distant Universe (galaxies or the CMB). If the new redshift is not subtracted in the calculation for the cosmic expansion rate, the calculated value would be larger than the actual one. The Hubble constant tension as well as the problem of the cosmological constant is explained by this dynamical dark energy scenario. Since only the combinations of the model parameters appear in the equations of cosmic evolution, the separate values of the parameters cannot be determined completely by astronomical observations based solely on the homogeneous and isotropic cosmology. In order to break the parametric degeneracy, we need to analyze the data of astronomical observations and laboratory experiments in the case of non-uniform distribution of matter, because the three-dimensional force mediated by a scalar field depends on its spatial gradient.

\begin{acknowledgments}
I acknowledge discussions with Jie-Nian Zhang, Zhao-Hong Li, Trent Claybaugh and Chuan Wang. This work was supported by the National Natural Science Foundation of China through Grant No. 12074396 and Science Challenge Project through Grant No. TZ2018003.
\end{acknowledgments}

\appendix

\section{The detailed derivation of the frequency shift of light signal}\label{appendixA}

The Friedmann-Roberson-Walker (FRW) metrics are
\begin{equation}\label{3pr221207}
{g_{\mu \nu }} =
\left( {\begin{array}{*{20}{c}}
  { - 1}&{}&{}&0 \\
  {}&{\frac{{{a^2}}}{{1 - K{r^2}}}}&{}&{} \\
  {}&{}&{{a^2}{r^2}}&{} \\
  0&{}&{}&{{a^2}{r^2}{{\sin }^2}\theta }
\end{array}} \right)
\end{equation}
and
\begin{equation}\label{3pr2212072}
{g^{\mu \nu }} = \left( {\begin{array}{*{20}{c}}
  { - 1}&{}&{}&0 \\
  {}&{\frac{{1 - K{r^2}}}{{{a^2}}}}&{}&{} \\
  {}&{}&{\frac{1}{{{a^2}{r^2}}}}&{} \\
  0&{}&{}&{\frac{1}{{{a^2}{r^2}{{\sin }^2}\theta }}}
\end{array}} \right),
\end{equation}
respectively. Consequently, the nonzero connection coefficients can be calculated, and can also be found in popular textbooks of general relativity, as follows:
\begin{equation}\label{3pr2212073}
\begin{gathered}
\Gamma _{11}^0 = \frac{{{a^2}\dot a}}{{(1 - K{r^2})a}} = H{g_{11}},\Gamma _{22}^0 = a\dot a{r^2} = H{g_{22}},\Gamma _{33}^0 = a\dot a{r^2}{\sin ^2}\theta  = H{g_{33}},  \hfill \\
  \Gamma _{10}^1 = \Gamma _{20}^2 = \Gamma _{30}^3 = H,\Gamma _{11}^1 = \frac{{Kr}}{{1 - K{r^2}}},\Gamma _{22}^1 =  - r(1 - K{r^2}),\Gamma _{33}^1 =  - r(1 - K{r^2}){\sin ^2}\theta , \hfill \\
  \Gamma _{12}^2 = \frac{1}{r},\Gamma _{33}^2 =  - \sin \theta \cos \theta ,\Gamma _{13}^3 = \frac{1}{r},\Gamma _{23}^3 = \cot \theta, \hfill \\
\end{gathered}
\end{equation}
where $H=\dot{a}/a$ denotes the Hubble parameter. In order to obtain the relation Eq. (\ref{3prl1123}), we should solve the equations of motion for light signal as shown by Eqs. (\ref{scicence11093}), for $k^0$ component of the four-wave-vector, one has
\begin{equation}\label{3pr2212075}
{\left[ {\left( {1 + \mathcal{B}} \right)k^0} \right]_{;\nu }}\frac{{d{x^\nu }}}{{d\sigma }} = \left( {1 + \mathcal{B}} \right){k^{0}_{,\nu }}\frac{{d{x^\nu }}}{{d\sigma }} + \left( {1 + \mathcal{B}} \right)\Gamma _{\lambda \nu }^0 {k^\lambda }\frac{{d{x^\nu }}}{{d\sigma }} + {\left( {1 + \mathcal{B}} \right)_{,\nu }}\frac{{d{x^\nu }}}{{d\sigma }}{k^0}.
\end{equation}
Inserting Eqs. (\ref{3pr2212073}) into Eq. (\ref{3pr2212075}), one has
\begin{equation}\label{3pr2212076}
\frac{{d{k^0}}}{{d\sigma }} + \frac{{d\ln \left( {1 + \mathcal{B}} \right)}}{{d\sigma }}{k^0} + H{g_{11}}{k^1}\frac{{d{x^1}}}{{d\sigma }} + H{g_{22}}{k^2}\frac{{d{x^2}}}{{d\sigma }} + H{g_{33}}{k^3}\frac{{d{x^3}}}{{d\sigma }} = 0.
\end{equation}
Inserting ${k_1} = {g_{1\nu }}{k^\nu } = {g_{11}}{k^1}, {k_2} = {g_{22}}{k^2}$ and ${k_3} = {g_{33}}{k^3}$ into above expression, one has
\begin{equation}\label{3pr230203}
\frac{{d{k^0}}}{{d\sigma }} + \frac{{d\ln \left( {1 + \mathcal{B}} \right)}}{{d\sigma }}{k^0} + H{k_1}\frac{{d{x^1}}}{{d\sigma }} + H{k_2}\frac{{d{x^2}}}{{d\sigma }} + H{k_3}\frac{{d{x^3}}}{{d\sigma }} = 0.
\end{equation}
Noticing the relation (\ref{scicence11156}) in the main text, i.e.,
\begin{equation}\label{3pr2212078}
 {k_0}\frac{{d{x^0}}}{{d\sigma }} =  - {k_i}\frac{{d{x^i}}}{{d\sigma }}
\end{equation}
with $i=1,2,3$, one has
\begin{equation}\label{3pr2212079}
\frac{{d{k^0}}}{{d\sigma }} + \frac{{d\ln \left( {1 + \mathcal{B}} \right)}}{{d\sigma }}{k^0} + H( - {k_0}\frac{{d{x^0}}}{{d\sigma }}) = 0.
\end{equation}
Since ${k_0} = {g_{0\nu }}{k^\nu } = {g_{00}}{k^0} =  - {k^0}$, one has
\begin{equation}\label{3pr2302032}
\frac{{d{k^0}}}{{d\sigma }} + \frac{{d\ln \left( {1 + \mathcal{B}} \right)}}{{d\sigma }}{k^0} + H{k^0}\frac{{d{x^0}}}{{d\sigma }} = 0.
\end{equation}
Consequently, we can choose the curve parameter
\begin{equation}\label{3pr22120710}
\sigma=x^0=t.
\end{equation}
Noticing $k^0$ is related to the frequency $\omega$ of the electromagnetic wave by the relation
\begin{equation}\label{3pr22120711}
k^0=\omega,
\end{equation}
Eq. (\ref{3pr2212079}) becomes as
\begin{equation}
\frac{{d\omega }}{{dt}}+ \left( { H + \frac{1}{{ {1 + \mathcal{B}} }}\frac{{d\mathcal{B}}}{{dt}}} \right)\omega =0. \label{3pr22120712}
\end{equation}


\begin{thebibliography}{100}
\bibitem{3prl1122} Based on the principle of equivalence, from a purely mathematical point of view, the mass parameter $M $ is unnecessary for Eq. (\ref{3prd1115}). The difference between photons and massive particles is mainly reflected by whether the line element $ds $ is zero or not.
\bibitem{3prdnote} The operator is defined by ${\nabla^2_{\rm{FRW}}} = \frac{{\sqrt {1 - K{r^2}} }}{{{r^2}}}\frac{\partial }{{\partial r}}\left( {{r^2}\sqrt {1 - K{r^2}} \frac{\partial }{{\partial r}}} \right) + \frac{1}{{{r^2}\sin \theta }}\frac{\partial }{{\partial \theta }}\left( {\sin \theta \frac{\partial }{{\partial \theta }}} \right) + \frac{1}{{{r^2}{{\sin }^2}\theta }}\frac{{{\partial ^2}}}{{\partial {\varphi ^2}}}$.
\bibitem{note1} In a contracting universe of $H<0$, the amplitude of the negative-damping oscillation will increase with time, the description of the cosmological constant will gradually become invalid with the passage of time.
\end{thebibliography}
\end{document}